\documentclass[pre,floatfix,twocolumn]{revtex4}  

\usepackage{epsfig}
\usepackage{graphicx}
\usepackage{amsmath}
\usepackage{ulem} 

\begin{document}

\author{Emily A. Hobbs, Alexander Christensen}
\author{Brian C. Utter}
\email{brian.utter@bucknell.edu}
\affiliation{Department of Physics and Astronomy, Bucknell University, Lewisburg, PA 17837, USA}
\date{\today}
\title{Clogging in bidirectional suspension flow}

\renewcommand{\textfraction}{0.05}
\renewcommand{\topfraction}{0.95}
\renewcommand{\bottomfraction}{0.95}
\setcounter{bottomnumber}{4} 
\setcounter{topnumber}{4}
\renewcommand{\floatpagefraction}{0.95}

\newcommand{\ea}{{\it et al.} }
\newcommand{\arrow}{$\longrightarrow$ }
\newcommand{\bcite}{{\it \bf [CITE]}}
\newcommand{\bcheck}{{\it \bf [CHECK]}}

\begin{abstract}        
The sudden arrest of motion due to confinement is commonly observed via the clogging transition in the flow of particles through a constriction. We present results of a simple experiment to elucidate a similar transition in the bidirectional flow of two species in which two species of macroscopic particles with different densities are confined in a tube and suspended in a fluid of intermediate density. Counterflowing grains serve as mobile obstacles and clogging occurs without arch formation due to confinement. We measure the clogging or jamming probability $J$ as a function of number of particles of each species $N$ in a fixed channel length for channel widths $D = $ 3$-$7$d$, where $d$ is the particle diameter. $J(N)$ exhibits a sigmoidal dependence and collapses on a single curve $J(N/D^3)$ indicating the transition occurs at a critical density. Data is well-fit by a probabilistic model motivated by prior constriction flows which assumes grains enter the clogging region with a fixed probability to produce a clogging state. A quasi-two-dimensional experiment  provides insight into the interface shape and and we identify a Rayleigh-Taylor instability at large channel widths. 
\end{abstract}


\maketitle



\section{Article}

When particle flow is confined by local geometry, a transition to clogging can be observed at sufficiently high particle density. In the prototypical example of gravity-driven grains falling through an opening in a hopper, an arch might spontaneously form across the orifice, supporting the weight of the grains above and leading to a blockage of flow. This transition is generic across a wide range of systems, including granular materials, colloids, and pedestrian traffic, and can be characterized by a clogging phase diagram in which increased particle density, increased compatible loading (such as  confining pressure stabilizing an arch of grains), or decreased incompatible load (such as fluctuations induced by ambient vibration) promote clogging \cite{Zuriguel:14:Clogging}. Such systems exhibit common statistical features including power-law distributions of time between consecutive particles and exponential distributions for the size of particle bursts between clogging events  \cite{Zuriguel:14:Clogging,Zuriguel:03:Jamming,Zuriguel:05:Jamming,To:05:Jamming}, which suggests a  
constant probability of clogging during flow \cite{Zuriguel:03:Jamming,Arevalo:16:Clogging}.

As particles are driven through an orifice much larger than the size of the particle, they typically flow at a constant rate \cite{Beverloo:61:Flow}. As the opening size $D$ decreases, the probability to clog increases in a sigmoidal curve rising from 0 to 1 over an opening of size $D \approx$ 2$-$5 $d$, where $d$ is the particle diameter \cite{To:01:Jamming,Zuriguel:03:Jamming,Janda:08:Jamming}.
As $D$ decreases, the lifetime of the flowing state before particle arrest decreases. 
When a finite number of particles $N$ is used, the jamming probability increases with $N$ for fixed $D$ \cite{Zuriguel:03:Jamming,Janda:08:Jamming} and the transition becomes sharper \cite{Janda:08:Jamming}. 
This clogging probability has historically been called the jamming probability $J$, though the system spanning arrest in jamming is an increasingly well characterized transition \cite{Liu:10:Jamming,Behringer:19:Physics} distinct from the arrest due to local geometry in clogging \cite{Peter:18:Crossover}. 
Similar behavior is observed in 
suspensions in which a wider range of particle concentrations and velocities are accessible  
and hydrodynamic effects may be relevant \cite{Guariguata:12:Jamming,Lafond:13:Orifice}. 

In seminal work, To \ea describe two-dimensional hopper flow of monodisperse disks  as a probabilistic process in which grains sample different configurations until, by chance, an arrangement of particles corresponding to a stable arch spans the orifice \cite{To:01:Jamming}. They find $J(D)$ agrees with a model based on a restricted random walker. 
Subsequent work by a variety of authors 
similarly model clogging consistent with the assumption that configurations of particles are sampled statistically independently until a stable configuration is reached 
\cite{Zuriguel:03:Jamming,To:05:Jamming,Janda:08:Jamming,Guariguata:12:Jamming,Lafond:13:Orifice,Thomas:15:Fraction}.
An exponential distribution of flow durations  implies clogging is a Poisson process where there is some large probability to remain unclogged at each time step. In hopper flows, it's assumed a new configuration occurs after a grain falls approximately a distance equal to its diameter \cite{Thomas:15:Fraction}. 

These probabilistic models typically assume that 
an individual grain will fall through the orifice with a large probability $p_1$ and the probability for a clog to form 
at the $n^\textrm{th}$ configuration is  $(1-p_1)p_1^{n}$ \cite{Janda:08:Jamming}. An initial transient of $n_T$ grains may occur before a steady-state concentration of uncorrelated states is reached
\cite{To:05:Jamming,Guariguata:12:Jamming,Lafond:13:Orifice}. However, 
the jamming probability per particle reaches a constant value and 
this transient can often be ignored \cite{Guariguata:12:Jamming,Thomas:15:Fraction} or estimated using data \cite{Lafond:13:Orifice}. 
For openings larger than $3d$, the probability that an individual grain will pass the orifice without leading to a clog is close to one \cite{Zuriguel:05:Jamming},
 related to the mean avalanche size \cite{Janda:08:Jamming}, and only weakly dependent on velocity \cite{Guariguata:12:Jamming} and driving force \cite{Arevalo:16:Clogging}.
The fraction of possible flowing grain configurations that precede a clog can be determined based on the average mass discharged before clogging \cite{Thomas:15:Fraction}.
If there are a fixed number of grains $N$ in the experiment, the probability to clog during the run is then the cumulative probability for $n < N$.

Dependence on dimension is less clear. A simple, probabilistic model based on arch formation predicts 
$J_N(D) = 1 - exp[-NAe^{-B(\eta_0 D)^2}]$ for 2d systems \cite{To:01:Jamming,Janda:08:Jamming}, where $\eta_0$ is indicates how the number of grains in an arch scales with $D$.
Janda \ea find using $D^3$ in the exponential for 3d as one might guess is not satisfactory \cite{Janda:08:Jamming}, though
Thomas and Durian take the number of grains in the clogging region to be $(D/d)^\alpha$ 
with $\alpha \approx 3$,
suggesting the volume of grains is the relevant factor \cite{Thomas:15:Fraction}. 
Avalanche size has been found to depend on $D^2$ in 2d simulations, which is the scale of the number of particles in the vicinity of the constriction \cite{Arevalo:16:Clogging}, and the opening area rather than the volume in 3d suspension clogging \cite{Lafond:13:Orifice}.

While much has been learned about clogging through orifices, relatively little is known about such behavior in bidirectional flow, in which two species attempt to pass each other as they are driven in opposite directions through a channel.
Despite its simplicity, the dynamics are intriguing due to nonlinear feedback as each particle species serves as mobile obstacles for the other.
In the absence of a constriction, clogging may still occur due to confinement, but 
stable configurations for bidirectional flow can not generically be arches, the basis of hopper clogs. 

A substantial portion of the work on bidirectional flow are via simulations, for instance 
models of pedestrian traffic \cite{Muramatsu:99:Jamming,Tajima:02:Pattern,Nowak:12:Quantitative}, typically as cellular automota or biased random walkers which may include a variety of social interactions between walkers, such as following or avoidance behavior.  
This work has characterized the phase diagram \cite{Nowak:12:Quantitative} and the so-called fundamental diagram characterizing flow rate versus density \cite{Zhang:12:Ordering,Flotterod:15:Bidirectional}. Brownian dynamics simulations have also been performed to model damped colloidal particles \cite{Dzubiella:02:Lane,Glanz:16:Symmetry} and cat-anionic lipid layers \cite{Netz:03:Conduction} in which two oppositely charged species are driven in an electric field as well as bidirectional flows of deformable short chains \cite{Mashiko:16:Flow}. 
There has been limited experimental work for  counterflow in both pedestrian \cite{Isobe:04:Experiment,Helbing:05:Self,Kretz:06:Experimental,Zhang:12:Ordering} 
and colloidal \cite{Vissers:11:Band,Vissers:11:Lane} systems.


A jamming transition is observed in bidirectional flow simulations, occurring at a critical density independent of system size, for instance for different width channels  \cite{Tajima:02:Pattern}. The transition density 
decreases with increasing drift speed \cite{Muramatsu:99:Jamming} 
and increases with the inclusion of social forces \cite{Nowak:12:Quantitative}
but is 
relatively insensitive to drift speed when only avoidance is included \cite{Tajima:02:Pattern}. 
The jamming probability increases with increasing density and monotonically increases with channel length to width ratio \cite{Nowak:12:Quantitative}.
If the interface is not flat, the lateral imbalance of particles can lead to particles pushing through to break the clog \cite{Helbing:00:Freezing}.
This is reminiscent of a Rayleigh-Taylor instability, observed in systems with a 
a density inversion, in with denser particles above a lower density layer \cite{Vinningland:07:Granular}.

\begin{figure}
\includegraphics[width=1.75in]{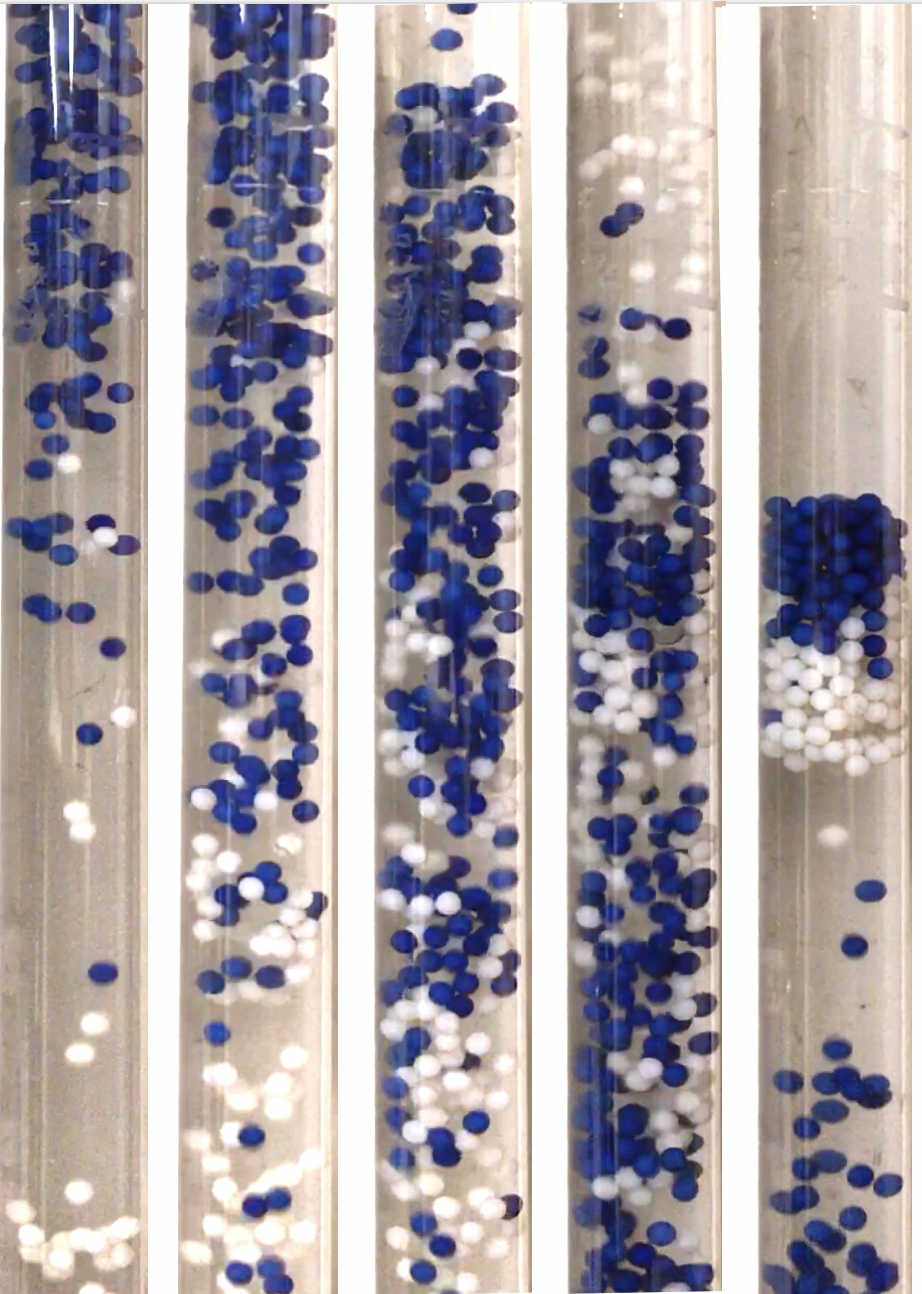}
\caption{Sequential images during the formation of a clog. The time interval is 3 s between images and $N = 200$. 
}
\label{SequentialImages2}
\end{figure}

In this article, we present experiments to quantify the clogging probability in bidirectional flow in which particles of two different densities in an intermediate density fluid are driven in opposite directions. Starting from opposite sides of a tube of constant diameter, they may pass through each other or form a static clog, as shown in Figure~\ref{SequentialImages2}. 
As noted above, there are important differences compared to hopper flow, namely 
(\textit{i}) clogging can occur due to confinement in a channel of constant width in the absence of a constriction, 
(\textit{ii}) the obstacles are themselves transient and continuously evolve until arrest, and 
(\textit{iii}) the stable geometry of a clog, by necessity, can not be an arch from the perspective of both species. 

The apparatus consists of macroscopic nylon and high-density polyethylene (HDPE), spherical beads of diameter $d$ =  6.4 mm (0.25") in a circular tube of diameter $D =$ 3$-$7 $d$ and length 1 m ($\approx$ 160 $d$) filled with a water/glycerol mixture. The nylon and HDPE spheres are monodisperse to within 0.4\% and 0.8\% respectively and the nylon spheres are dyed to visually distinguish the two types of particles. We use a glycerol concentration of 14\% by volume to produce a fluid density of $\rho = 1.04$ g/cm$^3$, intermediate between the densities of HDPE and nylon of $\rho = 0.94$ and 1.14 g/cm$^3$ respectively. The fluid is Newtonian with a viscosity approximately 1.8 times that of water. 

A particular number $N$ of each species is enclosed in the tube which is then mounted to a rotating armature. With the light/heavy particles initially at the top/bottom of the tube, the armature is quickly flipped to the opposite orientation. The tube is held vertical to within 0.5 degrees, as misalignment leads to the two species preferentially segregating laterally, reducing the clogging probability significantly.


During an experimental run, grains on each end disperse into a cloud of particles. 
Individual grains rapidly reach terminal velocity $v \approx 8$ cm/s. Grains interact through effectively inelastic collisions before sliding or rolling past each other. 
Upon reaching the interaction region, collisions can lead to substantial slowing of opposing particles. A clog forms if particles reach a mechanically stable arrangement. Visually, it is unclear until the last moment whether a clog will form or whether particles will cascade through each other. 
Fluctuations due to fluid effects are evident as spheres travel through the tube; despite the existence of a nonlinear and long-range interaction though, what follows is consistent, to lowest order, with a picture based on geometrical confinement and hard sphere interactions. This may be due to the substantial slowing that occurs at high density when clogging becomes likely. 
Friction is small, but not negligible, as slight asymmetries in the number of each particle may be stabilized by friction with the side wall.

\begin{figure}
\includegraphics[width=3.45in]{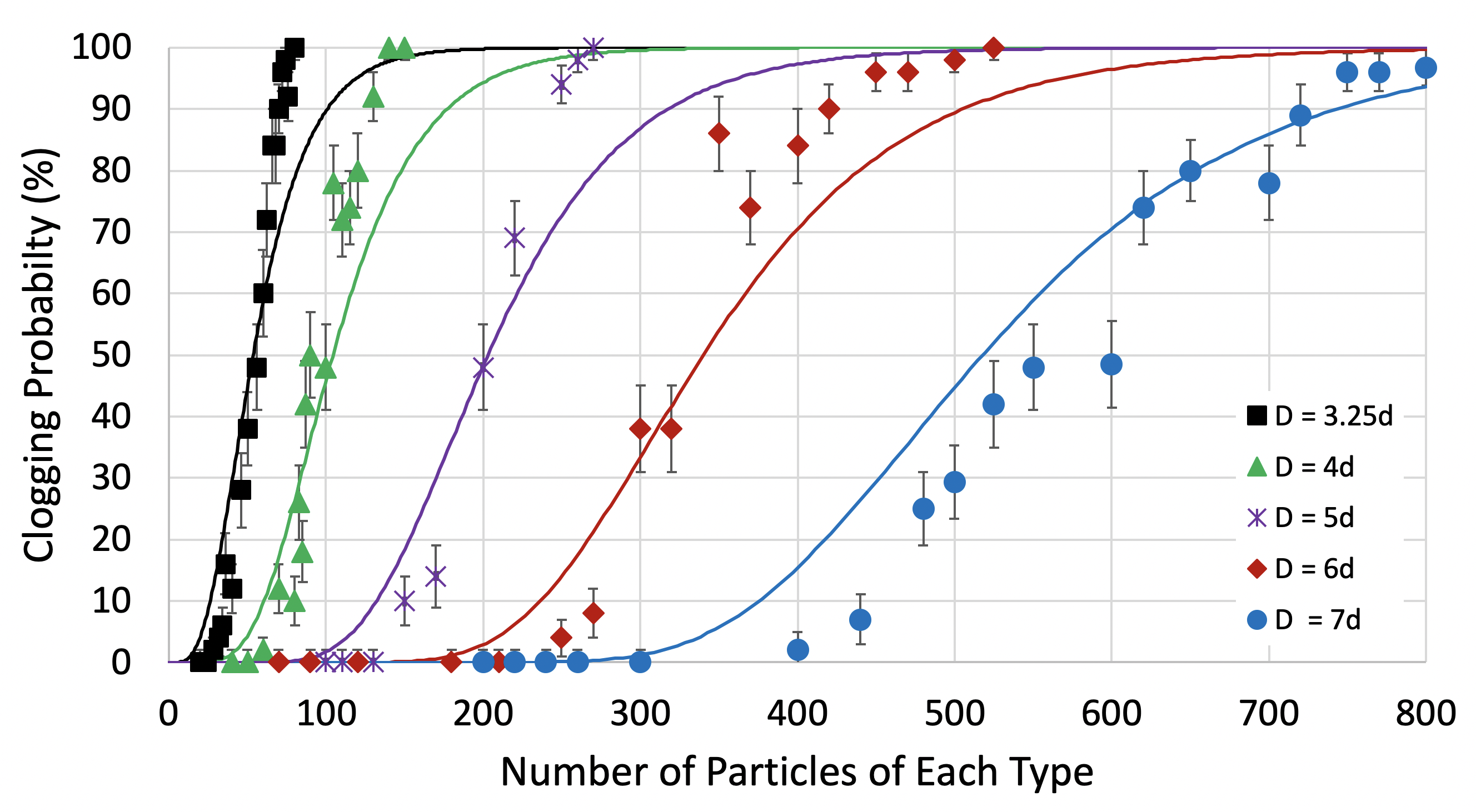}
\caption{Clogging probability for bidirectional flow of $N$ particles of each species. Different curves represent data from different diameter tubes $D$, increasing from $3d$ to $7d$ from left to right. The lines represent results of a model described in the text. 
}
\label{CloggingProbability}
\end{figure}

We measure the clogging or jamming probability $J(N)$ as the fraction of runs leading to a static clog, typically for 50-100 attempts per data point, as a function of number of beads $N$ of each type for tube diameter $D$ 
as shown by the data points in
Fig.~\ref{CloggingProbability}. We observe a sigmoidal probability distribution $J(N)$ with probability varying from 0 to 100\% over a relatively narrow range of particle number, reminiscent of jamming probability versus opening size for hopper flow \cite{To:01:Jamming,Zuriguel:03:Jamming,Janda:08:Jamming}. 
The probability to clog increases rapidly with number of particles as larger numbers of grains encounter larger numbers of obstacles impeding their flow.
For larger tube diameters, the curve shifts to the right towards larger particle number and the clogging transition broadens.

\begin{figure}
\includegraphics[width=3.45in]{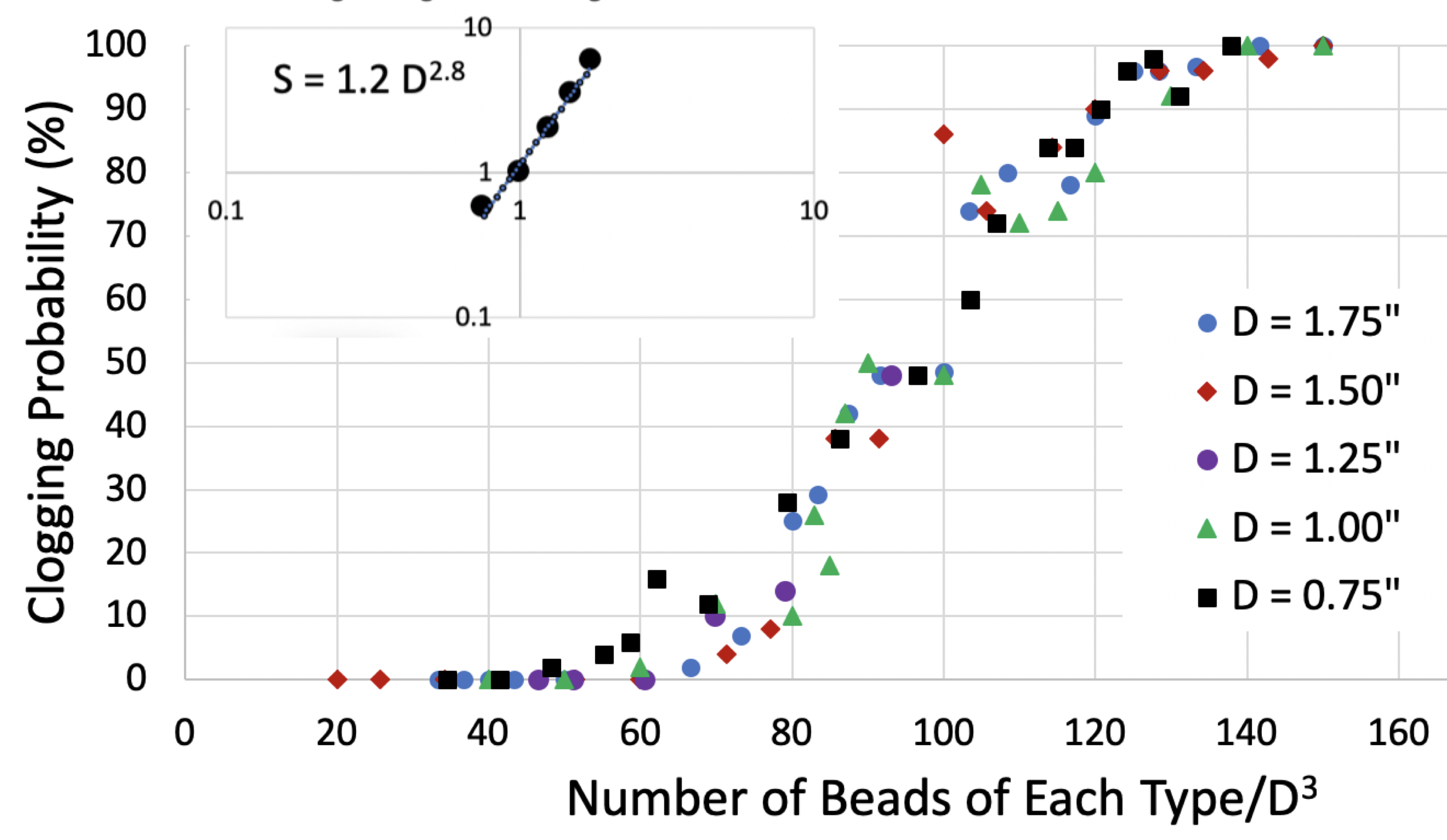}
\caption{Clogging probability versus number of beads divided by tube diameter cubed, $N/D^3$. (inset) The scaling factor $S$ leading to the best collapse of $J(N/S)$ is plotted versus tube diameter $D$. The exponent of 2.8 suggests $S \approx D^3$. 
}
\label{CloggingProbabilityRescaledD3}
\end{figure}

To collapse data onto a single curve, we rescale the horizontal axis by $D^3$
in Fig.~\ref{CloggingProbabilityRescaledD3}, indicating that to lowest order the transition depends on a particle density given by $N/D^3$ and that the clogging transition occurs when a critical density is reached.
Though the present experiment does not allow measurement of local packing fraction, 
grains spread out to extend a length of around 60 $d$ in the tube as they pass through each other corresponding to a packing fraction of order $\phi \approx 0.1$.  
In Fig.~\ref{CloggingProbabilityRescaledD3} (inset), we plot the scaling factor $S$ required to achieve the best data collapse for $J(N/S)$; the measured scaling exponent of $2.8 \approx 3$ is consistent with the cubic dependence $D^3$ we use in Fig.~\ref{CloggingProbabilityRescaledD3}.  



As the number of particles increases, the probability to form a clog at multiple locations within the tube also increases. 
It is known that the passing time, the time for all participants to pass a specific location, increases linearly with group size \cite{Kretz:06:Experimental}, leading to the possibility of an extended interaction region. The probability to form two clogs increases in a similar sigmoidal curve, beginning to rise when the single clog probability is approximately 50\%. We similarly observe the onset of three distinct clogs within the tube as the probability for two clogs becomes appreciable. 


The fact that  clogging displays a sigmoidal probability as in orifice flow suggests that a probabilistic explanation might similarly be employed. 
We  propose a simple model to ascertain whether a probabilistic approach as used previously for hopper flow may also be appropriate for bidirectional flow. 
We define 
the approximate number of grains that fit in the cylindrical clog region 
given a packing fraction $\phi_0$ of approaching grains as
$N_c =  \phi_0 D \pi \left(\tfrac{D}{2}\right)^2 / \tfrac{4}{3}\pi \left(\tfrac{d}{2}\right)^3 = 
\tfrac{3}{2} \phi_0 \tfrac{D^3}{d^3}$ . Empirically,  a value $\phi_0 \approx 0.1$ produces the best fit, consistent with the estimate above based on experimental images.
There is a new configuration after some characteristic time $\tau$, 
which we number with integer $n$.
At each configuration, we assume a probability $p_0$ that the configuration of incoming grains will not lead to a clog.
The probability that the grains might contribute to a clog at configuration $n$ after the prior $n$$-$1 non-clogging configurations is then $p_0^{n-1} (1-p_0)$.
We estimate the number of grains entering between configurations separated by this characteristic $\tau$ to be some fraction of the number of grains passing through the cross-sectional area of the tube, $N_0 = a_0 (D/d)^2$ such that $N = N_0 n$. We again find that $a_0 \approx  0.1$ leads to the best fit for this system, comparable to the packing fraction $\phi_0$. 
The total or cumulative probability that the grains will reach a clogged state by configuration $n$ is $p_t(n) = \sum_1^n   p_0^{n-1} (1-p_0)$.
The probability that the $N_c$ collection of grains are in a clogged configuration at the same time 
is $\left(p_t(n)\right)^{N_c}$ which
results in a final jamming probability 
\begin{equation}
J(n) = \left[ \sum_1^n   p_0^{n-1} (1-p_0) \right]^{N_c}
\end{equation}
This is plotted versus $N$ ($= N_0 n$, as defined above) as solid lines in Figure~\ref{CloggingProbability} with $p_0 = 0.96$,  consistent with similar models of hopper flows.
We note there are a number of simplifications assumed in this model, including the assumption of a constant clogging probability independent of tube diameter and neglecting the dependence on local particle density and fluid dynamics. However, there is a  single set of parameters for all fits in Fig.~\ref{CloggingProbability} which captures the general behavior suggesting that a probabilistic, geometric model may also be appropriate for bidirectional flow. 

We gain further insight by performing a second set of experiments in a quasi-two-dimensional channel in which 0.125" thick nylon and HDPE disks of diameter 5/16" (7.9 mm) are contained between two plexiglass sheets separated by a distance slightly larger than 0.125" to form a channel of length 70 $d$. The opening width $D$ can easily be increased beyond what is feasible in the 3d experiment to explore wider channels and we track all particles to study dynamics and characterize jamming interface. Detailed results will be the focus of a future study, but 
we gain insight into the shape of the interface and key differences compared to hopper flow.

\begin{figure}
\includegraphics[width=3.45in]{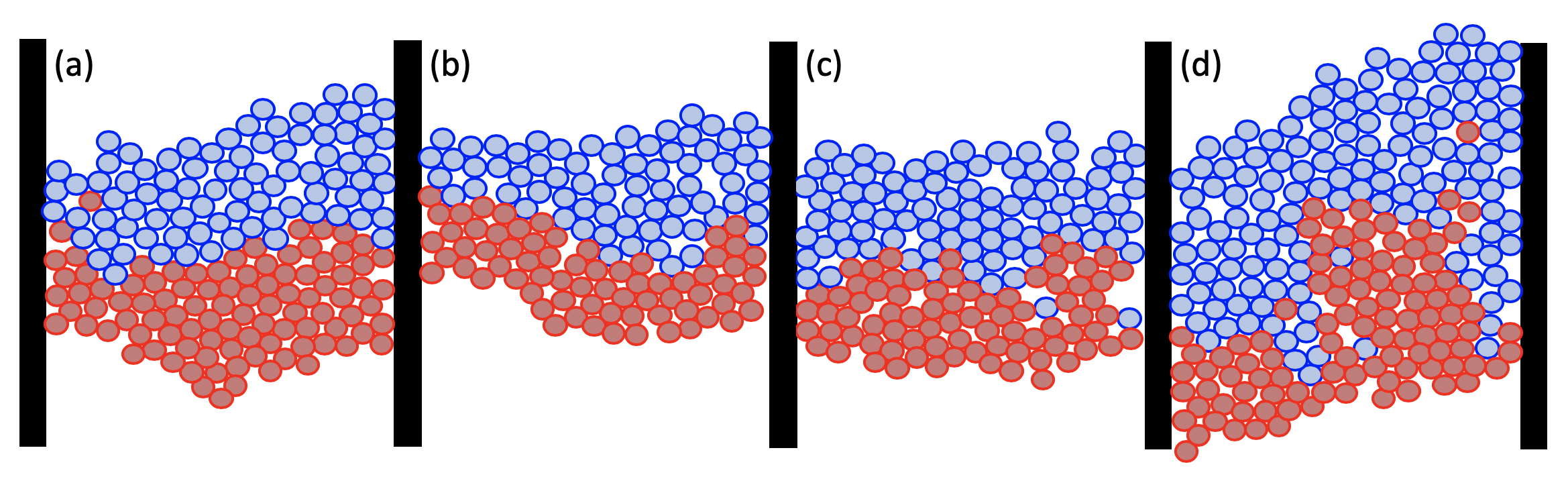}
\caption{Extracted profiles of static clogs from a quasi-two-dimensional experiment.  }
\label{Example2dClogs}
\end{figure}

\begin{figure}
\includegraphics[width=2.8in]{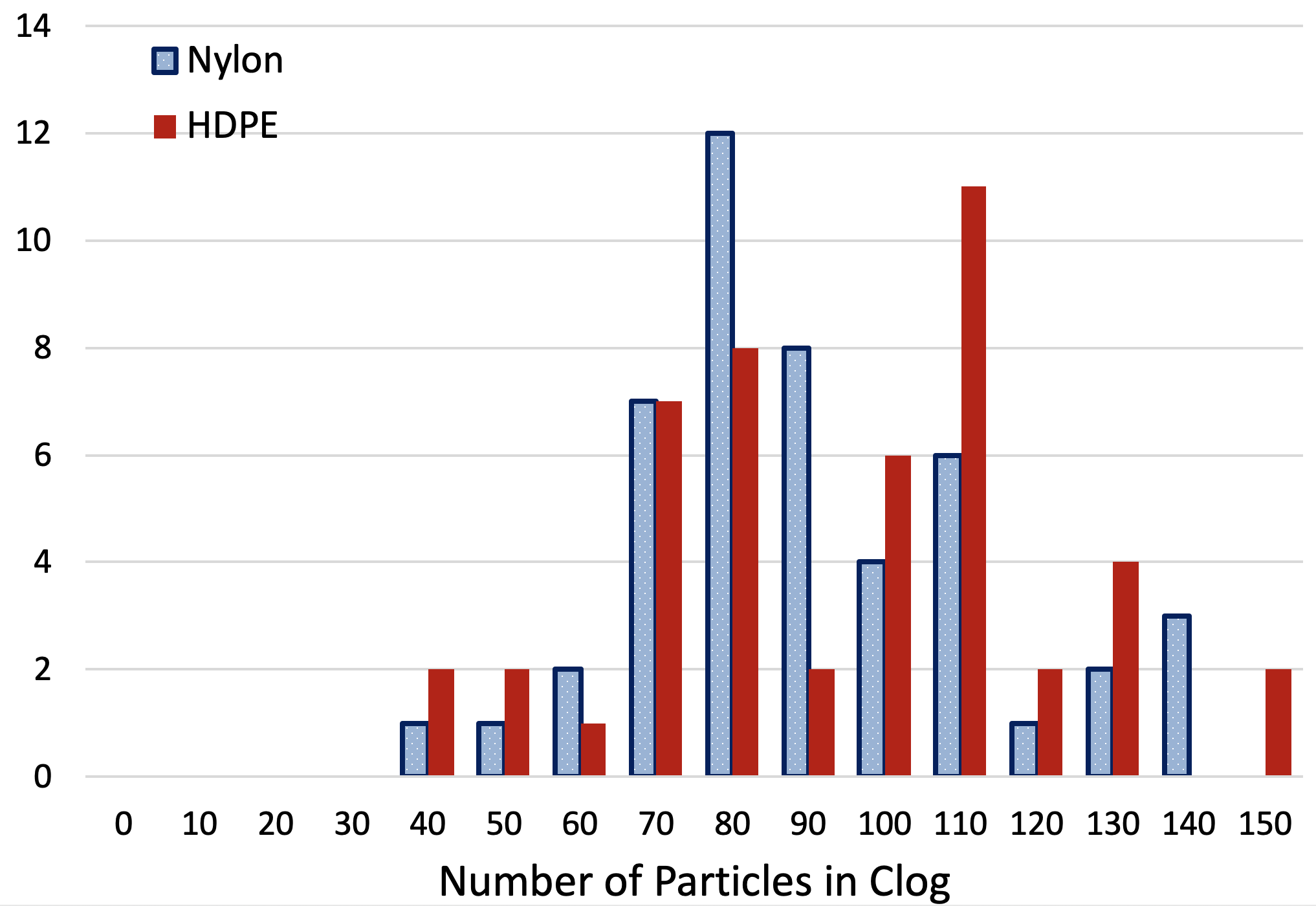}
\caption{Histogram of the number of particles of each type in a clog.}
\label{NumParticlesInJam2}
\end{figure}

In Fig.~\ref{Example2dClogs}, we show particle positions extracted from experimental clogs in the 2d geometry. We observe that clogs can form in much wider channels than observed in hopper flow (here around $16d$). We note that the interface of a clog is frequently fairly flat, as may be expected by the symmetry of the experiment and, as expected, is not comprised of arch-like structures. However, roughness at the grain scale is apparent and sometimes striking, including inclusions and plumes, seen in Figures~\ref{SequentialImages2} and~\ref{Example2dClogs}. Fig.~\ref{Example2dClogs}(d) hints at a failure mechanism we frequently observe in wider channels in which a large plume of falling particles on the left and rising particles on the right may push through in a Rayleigh-Taylor like instability  in wider channels as  the interface rotates and breaks. This likely leads to deviations from a simple probabilistic model and may represent the development of lane formation \cite{Dzubiella:02:Lane,Netz:03:Conduction}, in which at sufficient driving force opposing traffic forms lanes to minimize collisions, previously observed in both colloidal and pedestrian experiments \cite{Vissers:11:Band,Vissers:11:Lane,Kretz:06:Experimental,Zhang:12:Ordering}. 


The obstruction typically contains comparable numbers of each species as might be expected by symmetry of the experimental flow.
Fig.~\ref{NumParticlesInJam2} shows a plot of the number of grains of each species in a clog for 50 runs. A typical clog in an experiment with $N$ = 160 grains and $D \approx 16 d$ is $88 \pm 25$ on each side.
Imbalances can be stabilized by wall friction such that the distribution of the difference in nylon versus HDPE particles is comparable ($\sigma \approx 30$).


In summary, bidirectional flow is a remarkably simple geometry that exhibits clogging behavior due to confinement and in the absence of a constriction. 
Yet simple geometrical origins well-studied in hopper flow, namely arch formation, are not possible and 
the two species exhibit strongly nonlinear interactions as mobile obstacles for each other and mediated by fluid flows. This raises the question whether the clogging statistics and mechanism might be similar to those observed for particles flowing through a constriction. 
We measure a sigmoidal jamming probability $J(N)$ as a function of the number of each type of grain $N$.
Rescaling of data as $J(N/D^3)$ indicates that, to lowest order, the transition depends on reaching a critical density within the channel, independent of channel width for the values studied. 
A simple probabilistic model captures the general behavior of the system, suggesting that the relevant mechanism is likely based on randomly sampling of configurations until a stable assembly is reached, as previously determined for hopper flow. Preliminary experiments in two-dimensional channels indicate though that for wider channels, a Rayleigh-Taylor instability develops due to lateral variations in particle number. This may represent the development of lane formation and likely limits the applicability of a purely geometric, probabilistic model. Further studies are needed to understand the detailed relationship between particle dynamics during clogging and interface morphology.

\begin{acknowledgments}
The authors would like to thank A. Graves for helpful discussions and for performing the pedestrian simulations that prompted this work. This material is based upon work supported by the National Science Foundation under Grant No. 1317446. 
\end{acknowledgments}


\begin{thebibliography}{10}

\bibitem{Zuriguel:14:Clogging}
I. Zuriguel {\it et~al.}, {Scientific Reports} {\bf {4}},  7324  ({2014}).

\bibitem{Zuriguel:03:Jamming}
I. Zuriguel, L. A. Pugnaloni, A. Garcimartin, and D. Maza, {Phys. Rev. E} {\bf
  {68}},  030301(R)  ({2003}).

\bibitem{Zuriguel:05:Jamming}
I. Zuriguel, A. Garcimartin, D. Maza, L. A. Pugnaloni, and J.M. Pastor, {Phys. Rev. E} {\bf {71}},  051303  ({2005}).

\bibitem{To:05:Jamming}
K. To, {Phys. Rev. E} {\bf {71}},  060301(R)  ({2005}).

\bibitem{Arevalo:16:Clogging}
R. Arevalo and I. Zuriguel, {Soft Matter} {\bf {12}},  123  ({2016}).

\bibitem{Beverloo:61:Flow}
W.~A. Beverloo, H.~A. Leniger, and J. van~de Velde, {Chemical Engineering
  Science} {\bf {15}},  {260}  ({1961}).

\bibitem{To:01:Jamming}
K. To, P. Y. Lai, and H. K. Pak, {Phys. Rev. Lett.} {\bf {86}},  71  ({2001}).

\bibitem{Janda:08:Jamming}
A. Janda {\it et~al.}, {Europhysics Letters} {\bf {84}},  44002  ({2008}).

\bibitem{Liu:10:Jamming}
A.~J. Liu and S.~R. Nagel, Annual Review of Condensed Matter Physics {\bf 1},
  347  (2010).

\bibitem{Behringer:19:Physics}
R.~P. Behringer and B. Chakraborty, Reports on Progress in Physics {\bf 82},
  012601  (2019).

\bibitem{Peter:18:Crossover}
H. P\'eter, A. Lib\'al, C. Reichhardt, and C.~J.~O. Reichhardt, Scientific
  Reports {\bf 8},  10252  (2018).

\bibitem{Guariguata:12:Jamming}
A. Guariguata, M. A. Pascall,  M. W. Gilmer, A. K. Sum, E. D. Sloan, C. A. Koh, and D. T. Wu, Phys. Rev. E {\bf 86},  061311  (2012).

\bibitem{Lafond:13:Orifice}
P.~G. Lafond, M. W. Gilmer, C. A. Koh, E. D. Sloan, D. T. Wu, and A. K. Sum, {Phys. Rev. E} {\bf {87}},  042204  ({2013}).

\bibitem{Thomas:15:Fraction}
C.~C. Thomas and D.~J. Durian, {Phys. Rev. Lett.} {\bf {114}},  178001
  ({2015}).

\bibitem{Muramatsu:99:Jamming}
M. Muramatsu, T. Irie, and T. Nagatani, Physica A {\bf 267},  487  (1999).

\bibitem{Tajima:02:Pattern}
Y. Tajima, K. Takimoto, and T. Nagatani, {Physica A} {\bf {313}},  709
  ({2002}).

\bibitem{Nowak:12:Quantitative}
S. Nowak and A. Schadschneider, Phys. Rev. E {\bf 85},  066128  (2012).

\bibitem{Zhang:12:Ordering}
J. Zhang, W. Klingsch, A. Schadschneider, and A. Seyfried, J. Stat. Mech.
  P02002  ({2012}).

\bibitem{Flotterod:15:Bidirectional}
G. Flotterod and G. Laemmel, {Transportation Research B} {\bf {71}},  194
  ({2015}).

\bibitem{Dzubiella:02:Lane}
J. Dzubiella, G. P. Hoffmann, and H. Lowen, Phys. Rev. E {\bf 65},  021402
  (2002).

\bibitem{Glanz:16:Symmetry}
T. Glanz, R. Wittkowski, and H. Lowen, {Phys. Rev. E} {\bf {94}},  052606
  ({2016}).

\bibitem{Netz:03:Conduction}
R.~R. Netz, Europhys. Lett. {\bf 63},  616  (2003).

\bibitem{Mashiko:16:Flow}
T. Mashiko and T. Fujiwara, {Physics Letters A} {\bf {380}},  3490  ({2016}).

\bibitem{Isobe:04:Experiment}
M. Isobe, T. Adachi, and T. Nagatani, Physica A {\bf 336},  638  (2004).

\bibitem{Helbing:05:Self}
D. Helbing, L. Buzna, A. Johansson, and T. Werner, {Transportation Science}
  {\bf {39}},  1  ({2005}).

\bibitem{Kretz:06:Experimental}
T. Kretz {\it et~al.}, {J. Stat. Mech.}  P10001  ({2006}).

\bibitem{Vissers:11:Band}
T. Vissers, A. van Blaaderen, and A. Imhof, {Phys. Rev. Lett.} {\bf {106}},
  228303  ({2011}).

\bibitem{Vissers:11:Lane}
T. Vissers {\it et~al.}, Soft Matter {\bf 7},  2352  (2011).

\bibitem{Helbing:00:Freezing}
D. Helbing, I. J. Farkas, and T. Vicsek, {Phys. Rev. Lett.} {\bf {84}},  1240
  ({2000}).

\bibitem{Vinningland:07:Granular}
J.~L. Vinningland, O. Johnsen, E. G. Flekkoy, R. Toussaint, and K. J. Maloy, {Phys. Rev. Lett.} {\bf {99}},  048001
  ({2007}).

\end{thebibliography}

\end{document}